# Space-to-ground quantum-communication using an optical ground station: a feasibility study

P. Villoresi[a], F. Tamburini[b], M. Aspelmeyer[c], T. Jennewein[d], R. Ursin[c], C. Pernechele[e], G. Bianco[f], A. Zeilinger[c,d], C. Barbieri*[1][b1]

[a]Department of Engineering of Information, University of Padova, Italy and INFM-LUXOR, Padova Italy,
[b]Department of Astronomy, University of Padova, Italy,
[c]Institut für Experimentalphysik, Universität Wien, Austria,
[d]Institut für Quantenoptik und Quanteninformation, Österreichische Akad. der Wissenschaften, Wien, Austria,
[e]INAF Cagliari, Italy,
[f]Space Geodesy Center, ASI, Matera, Italy.

## ABSTRACT

We have tested the experimental prerequisites for a Space-to-Ground quantum communication link between satellites and an optical ground station. The feasibility of our ideas is being assessed using the facilities of the ASI Matera Laser Ranging Observatory (MLRO). Specific emphasis is put on the necessary technological modifications of the existing infrastructure to achieve single photon reception from an orbiting satellite.

## 1. INTRODUCTION

Quantum communications in space[1,2] has become a new technological challenge in the evolving field of quantum communications[3]. Its main goal is to achieve the distribution of single photons[4,5,6] or entangled photon pairs[7] from satellites to implement both quantum technologies such as quantum cryptography[8] and novel fundamental quantum physics experiments[9]. The main benefit of a space infrastructure is that it allows for much larger photon propagation distances than are achievable with earth bound quantum links, thus eventually enabling the build up of a global quantum communication network. Up to now, the status of such experiments is on the level of paper studies (see references above) and the availability of space qualified single photon sources is not yet foreseeable. However, to establish the necessary prerequisites for such experiments, it is important to investigate if and how present technologies of optical communication (i.e. satellite communication) can be utilized for quantum communication purposes. Specifically, we are interested in whether present optical ground stations can be modified to act as receiver stations to quantum communicate with future single-photon emitting satellites.

---

[1] cesare.barbieri@unipd.it; phone: +390498278221; fax +390498278245

## 2. SINGLE PHOTON QUANTUM COMMUNICATION

The underlying basis for quantum information-processing is the possibility to encode a logical bit onto a quantum system. Such qubits are subject to the fundamental laws of quantum physics (above all the superposition principle) and their use allows communication and computational tasks that are able to outperform their classical counterparts. To benefit from so-called quantum communication protocols such as quantum key distribution or quantum teleportation, it is mandatory to be able to distribute qubits between different communication partners. Since qubits are most easily implemented in single photons, the task is to transmit and receive single photons or, for more advanced schemes, entangled photon pairs. The most important applications would be quantum key distribution (QKD), which allows the generation of a secret key between distant parties inherently secured by the laws of quantum physics, and fundamental tests of quantum correlations over truly large distances. The theoretical properties of quantum information-processing and the feasibility of quantum communication in practical situations have been already elucidated by many experiments carried out in laboratories on the ground. In particular, recent advances in free-space experiments have allowed to achieve single-photon QKD over distances up to 23 km[10, 11] and the distribution of entangled photon pairs over 600 m (Aspelmeyer et al.[12]) using optical telescopes. In the long run, the application of space and astronomical technologies to quantum communication will enable further significant increases of these distances. The realization of a satellite-based quantum communication network communicating via ground stations is one of its attractive future perspectives.

## 3. AN OPTICAL GROUND STATION AS A QUANTUM RECEIVER

The conceptual scheme of a quantum communication receiver has already been outlined in Aspelmeyer et al.[1] and is reproduced in (Fig. 1). Following this idea, we intend to study the performance of an optical ground station for gathering ultra weak light signals while at the same time maintaining tracking of a satellite. These are two key features of a ground station used as a quantum communications receiver. To achieve that, we use weak laser pulses sent to and retroreflected down from a satellite using the Matera Laser Ranging Observatory (MLRO) in Matera, Italy.

In our experiment, the transmission and receiving line have the same optical paths, and use the polarization of light to discriminate the outgoing and incoming signals. The transmission channel optically couples a point source, realized by a small pinhole illuminated by a polarized laser beam at 532 nm, to the satellite via the telescope optics. The retroreflected photons are gathered by the telescope. In order to suppress the photons originated from different sources, as the celestial background, we have envisaged three different kinds of filtering: the spatial filtering, realized by a pinhole in the focal plane, a spectral filtering via a 0.15 nm bandpass interference filter (Andover, USA) and a selection of the recorded events according with the expected arrival time. This latter process requires the orbit reconstruction, as described below.

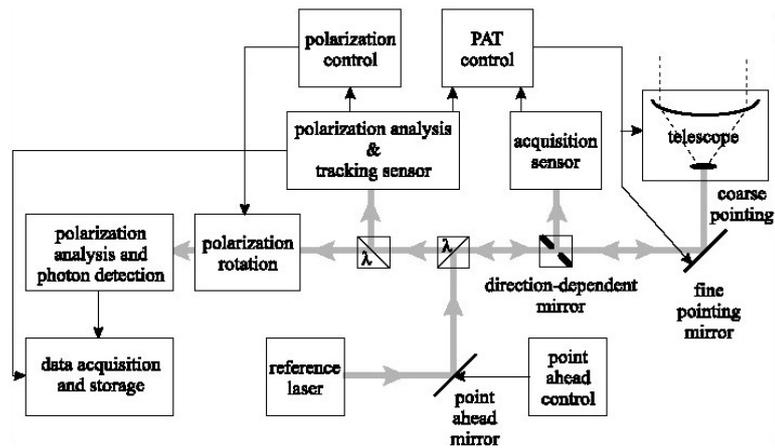

Fig. 1 –The conceptual scheme of a quantum communication receiver for photonic polarization qubits sent from a space based photon source (from Aspelmeyer et al.[1]).

The present paragraph explains the series of tests we are performing in order to demonstrate the feasibility of true quantum communication.

### 4.1 Description of MLRO

The MLRO (see Fig. 2) is a state of the art Satellite and Lunar Laser Ranging station, and is one of the best operational stations of the worldwide network managed by the International Laser Ranging Service ILRS (see[13] for a complete description of the ILRS network and activities).

The MLRO has three elements of great importance for Q-Space.

Firstly, it is equipped with a 1.5 meter primary mirror telescope with astronomical quality optics, and a long Coudè configuration f/212 realized by means of a small convex secondary mirrors plus 5 plane ones. While the primary mirror is wide-band Al-coated, all other mirrors are covered with a dielectric coating allowing excellent reflectivity in a narrow band pass centered around 532 nm. The beam divergence is diffraction limited and can be tuned in a continuous way from 1 to 20 arcsec.

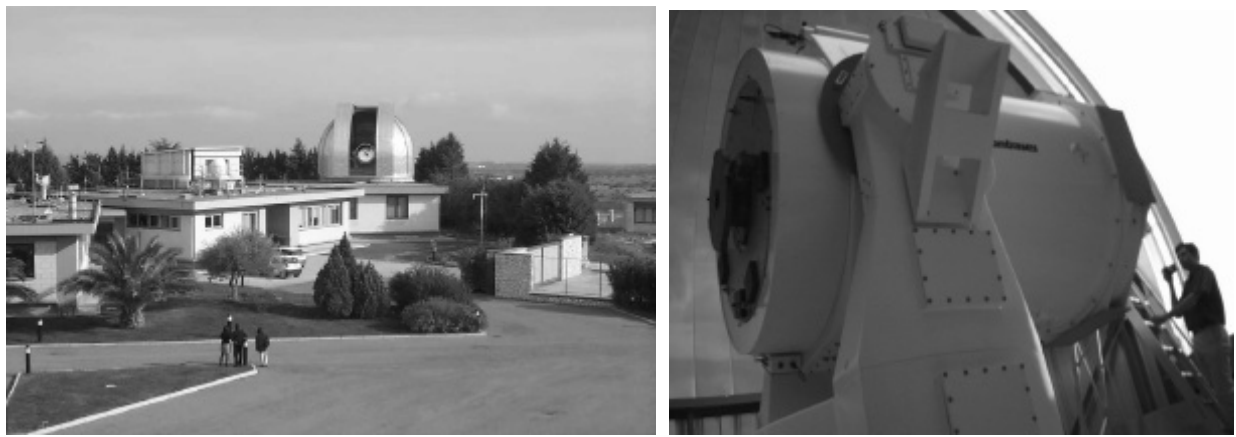

Fig. 2 – The Matera Laser Ranging Observatory (MLRO) and telescope

Second, the alt-azimuth mount is capable of a tracking velocity of 20 deg/sec in azimuth and 5 deg/sec in elevation, with a tracking accuracy of 1 arcsec RMS.

The third fundamental element of MLRO is the timing system, driven by a Cesium beam frequency standard, which is used to synchronize all our operations and time tag the photons to better than 0.01 nanoseconds.

MLRO regularly tracks all existing geodynamic satellites, in particular the Laser GEOdynamics Satellites **LAGEOS 1** (NASA) and **LAGEOS 2** (ASI). These very dense, passive, spherically shaped satellites, having a diameter of 60 cm and placed in a circular orbit with an height of approximately 5900 km, are equipped with 426 cube-corner retroreflectors built in such a way to send most of the signal back to the same direction of the incident light. Each retroreflector is essentially a cylindrical piece of fused silica having a 3.8cm in diameter and a 3-surface reflecting corner.

The MLRO laser is an active hybrid ND:YAG configured to emit 40ps long pulses in the wavelength of 532nm with an energy of 100mJ/pulse in the monochromatic setup. The averaged estimated number of photons per pulse is ~$5 \times 10^{17}$. The repetition rate of the laser system is 10 Hz, with a range timing accuracy of better than 2 ps.

**4.2 Tracking satellites with MLRO**

The conceptual scheme of the proposed experiment is as follows: the MLRO laser and telescope is initially used for approximately 1 to 3 minutes to acquire and track the satellite with high precision. Then, the dedicated transmitting laser, optical assembly and detector (in the following, the Quantum Transceiver QT) is inserted in the main optical transmission line for several more minutes. The cycle MLRO/QT is repeated for as many times as allowed by the satellite visibility (typically around 40 minutes in the case of Lageos, see Fig. 3).

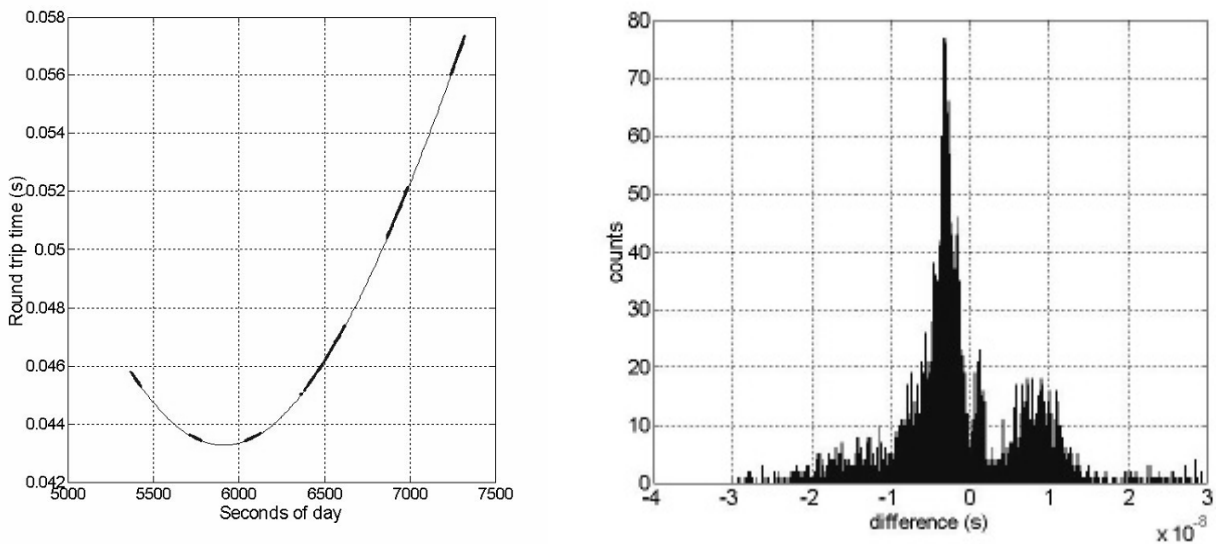

Fig. 3 – Scheme of satellite illumination with MLRO and QT. Left: time of flight to Lageos 1 vs time during the visibility. The dotted tracts indicate the time intervals when the MLRO laser is inserted to check the satellite pointing and tracking. Right: histogram of the differences between the measured data and the value predicted by polynomial fitting of order 60. .

This scheme has already been successfully tested. We recall that the time of flight of the photons retro-reflected by the satellite varies in time, due to the topocentric motion of the satellite itself. To determine the set of coincidences for the time-tagging, we perform a polynomial fit which is able to achieve an RMS difference between measured and fitted range points better than few nanoseconds (see again Fig.3). A better determination of the time of flight can be obtained *a posteriori* with the Geodyn II (NASA/GSFC) program[13], which fits the orbit through the observations with a typical post-fit residual RMS of about 15 mm in the position of the satellite. The MLRO post-fit shot-by-shot radial residuals for a well-observed pass are less than 1 cm RMS (often less than 5 mm RMS) for all targets.

### 4.3 – The photon link budget

A full understanding of the MLRO capabilities requires to take into account the full link budget (see Degnan[14]), which gives the number of reflected photons and then the number of photoelectrons, depending on different situations during the observation, as in the following equation:

$$N_{ph} = \eta_q E_T \frac{\lambda}{\hbar c} \eta_T G_T \sigma_{sat} \left( \frac{1}{4\pi R^2} \right)^2 A_T \eta_R T_A^2 T_c^2$$

where $\eta_q$ is the detector quantum efficiency, $\eta_T$ the transmitter path efficiency, $\eta_R$ the receiver path efficiency, $A_T$ the receiving aperture area of the telescope, $T_A$ the atmospheric transmission, $T_C$ the cloud transmission, $E_T$ the laser energy pulse, $\sigma_{sat}$ the satellite backscattering cross section and $G_T$ the transmitter gain. The link budget is function of $R^{-4}$, $R$ being the distance of the satellite from the observer (another way of expressing it is by using the square of the beam divergence $D$).

For the Lageos satellites, the main limiting factor of the efficiency is the size of the retroreflectors, which are only 3.8 cm in diameter. At the distance of the 6000 km, a beam divergence of 1 arcsec produces a spot 29 m wide: the energy fraction intercepted by the satellite would thus be $(0.6/29)^2 = 4.3 \times 10^{-4}$. This number is further reduced by one order of magnitude by taking into account the small number of active retroreflectors. The diffraction effects of the Lageos retroreflectors are of the order of 3", namely 100 m on the ground. The fraction intercepted by the MLRO telescope will thus be $(1.5/100)^2 = 2.3 \times 10^{-4}$. Furthermore, since the satellite is not stabilized, the retroreflected spot might have some additional beam spreading. The link budget can therefore be approximated to be about $10^{-10}$ as measured at the primary mirror of the MLRO telescope. A more conservative approach could lower this estimate by two orders of magnitude.

Other satellites at different heights and with larger retroreflectors give obviously different and more favorable values. On the other hand, the orbit of the Lageos is much more stable, and its visibility time is longer. Therefore, we plan to continue our tests with as many satellites as visible from the Matera station.

### 4.4 – Verification of photon efficiency by using stars

The Earth's turbulent atmosphere absorb a fraction of light and causes stars to "twinkle" with rapid intensity fluctuations. The measured probability distributions for the arrival of the photons in time arise from a combination of the atmospheric fluctuations with the Poisson distribution of the photon counts. Previous studies on the analysis of atmospheric intensity scintillation of stars[15] show that the time distribution of the photon counts is quite complicated, and it cannot easily be fitted neither with a Poissonian nor a Log-normal distribution. We have already done several star acquisitions with the MLRO and an APD detector to characterize the loss in transmission through the MLRO system, the efficiency of the detector, the effects of the seeing and the ability to time tag the detected photons.

We recall that the flux from a zero-magnitude star (like Vega) outside the atmosphere in the visual band V is given approximately by:

$$N_{ph} = 1 \times 10^3 \text{ cm}^{-2}\text{s}^{-1}\text{Å}^{-1}$$

Using the telescope effective area (1700 cm$^2$), the measured 7-mirror reflectivity (=0.9x0.97$^6$ = 0.70) over a bandwidth of 80 nm, the Avalanche Photodiode Quantum Efficiency and fiber coupling efficiency ( =0.1), the calculated transmissivity of the QT additional optical elements (= 0.29, including polarizer), an atmospheric absorption of approximately 0.3, the expected count rate from Vega is 1.5x10$^8$ s$^{-1}$. This number can be further reduced by the ratio between the pinhole area and the seeing FWHM in case of bad seeing. The unresolved sky background contributes a negligible signal.

Some experiments have already been performed. In the example shown in Fig.4, the counts from the bright star Vega were recorded for 91 seconds, and grouped in 0.01 sec wide bins. The power spectrum of the signal is shown in the right panel. A single narrow low frequency component is visible around 18 Hz and multiples. The origin of this spectral line is under investigation.

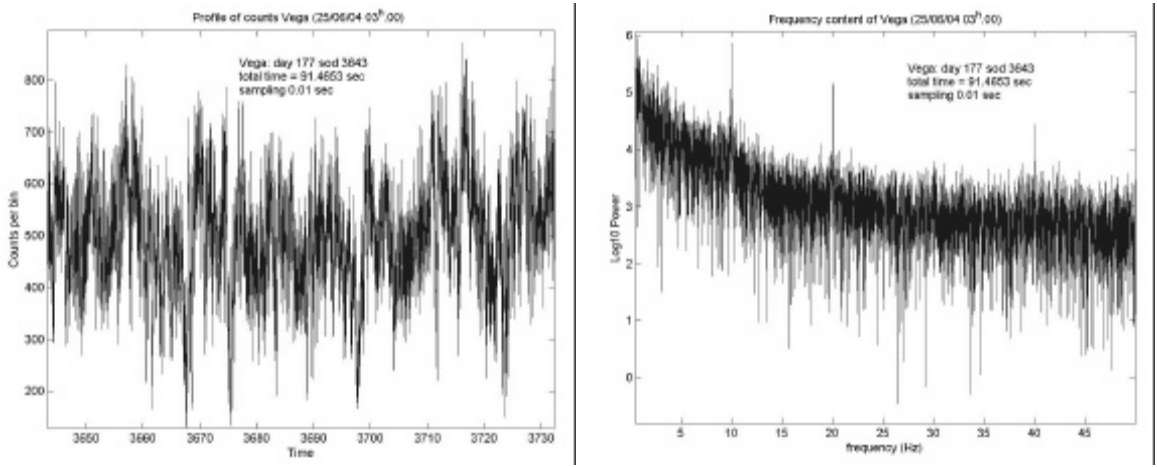

Fig. 4. Left: counts from Vega, binned over 0.01 s; right: power spectrum of Vega from 0 to 50 Hz.

**4.5 The alignment module**

In order to align our apparatus with the MLRO and determine the time scale offset due to the different cable lengths, detectors internal logic etc., we have devised an apparatus transmitting a laser beam with 17 kHz repetition rate toward the same calibration ground targets (retroreflectors located on the roofs of nearby buildings) used by MLRO.
The distances and two- way times of flight are respectively:
  Ground Target C : 45.25 m ,2x45.25 m $\Rightarrow$ 3.02x10$^2$ ns
  Ground Target B : 192.47 m, 2x192.47 m $\Rightarrow$ 1.28x10$^3$ ns

The offset has been determined equal to 116.1ns.
The alignment of QT with the main telescope is achieved when 100% of returns is obtained. Fig. 5 shows the histogram of the returned photons counts from 125000 pulses transmitted, reported in the time scale whose origin is the measure done by MLRO system on the same ground target.

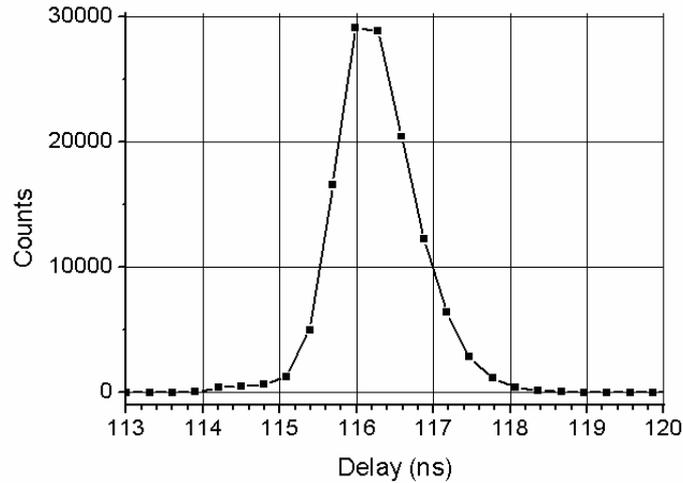

Fig. 5 - Histogram of the measured returns from a fixed target with respect to the time origin of the same measurement done by MLRO system. The actual setup shows an offset of about 116 ns and a FWHM time width of about 1.3 ns

The clear central peak contains all the returns; its width is caused by the pulse jitter and the electronic noise. The former is due to fluctuation in the laser pulses rep-rate, due to the pulse build-up mechanism in the Q-switching. The value of the peak width is very relevant in assessing the precision obtained by our system in the detection of the returned photons in the nanosecond scale.

## 5 - CONCLUSIONS

Our experiments already indicate the suitability of the MLRO telescope to act as a receiving station in a quantum communication experiment. This underlines our view that with existing technology the realization of a satellite-to-ground quantum communication link is actually feasible. Our work is intended to serve as the basis for future developments of dedicated systems for quantum communication between space and ground.

**Acknowledgments** – We wish to thank Drs. R. Sala and F. Schiavone, and the entire staff of MLRO Station for the very competent support and patient assistance. A. Gregnanin has greatly helped with the laboratory tests. This research has been substantially supported by the University of Padova under the Progetti di Ricerca di Ateneo 2003 program. We acknowledge further support by the Austrian Science Foundation (FWF) Project No. F1520, by the European Commission, Contract No. IST-2001-38864 (RamboQ), by the Alexander von Humboldt foundation and the Austrian Academy of Sciences.